\documentclass[twocolumn,trackchanges]{aastex61}
\usepackage{silence}
\usepackage{graphicx}
\graphicspath{{images/}}
\usepackage[caption=false]{subfig}
\usepackage{booktabs}
\usepackage{stmaryrd}

\usepackage{natbib}
\setcitestyle{authoryear,open={(},close={)}}

\usepackage{float}
\usepackage{color,soul}
\usepackage{wrapfig}
\usepackage{bm}
\usepackage{amsmath}

\usepackage{fancyhdr}
\usepackage{wasysym}

\newcommand{\sfrIR}{$\mathrm{SFR_{IR}}$}
\newcommand{\sfrUV}{$\mathrm{SFR_{UV}}$}
\newcommand{\mstar}{$\mathrm{M_*}$}
\newcommand{\mhalo}{$\mathrm{M_{halo}}$}
\newcommand{\lir}{$\mathrm{L_{IR}}$}
\newcommand{\alberts}{\citetalias{Alberts2021}}

\begin{document}

\defcitealias{Alberts2021}{A21}

\title{Measuring the total ultraviolet light from galaxy clusters at $z=0.5-1.6$: \\The balance of Obscured and Unobscured Star-Formation}

\author{Jed McKinney}
\affiliation{Department of Astronomy, University of Massachusetts, Amherst, MA 01003, USA}

\author{Vandana Ramakrishnan}
\affiliation{Department of Physics and Astronomy, Purdue University, 525 Northwestern Avenue, West Lafayette, IN 47907, USA}

\author{Kyoung-Soo Lee}
\affiliation{Department of Physics and Astronomy, Purdue University, 525 Northwestern Avenue, West Lafayette, IN 47907, USA}

\author{Alexandra Pope}
\affiliation{Department of Astronomy, University of Massachusetts, Amherst, MA 01003, USA}

\author{Stacey Alberts}
\affiliation{Steward Observatory, University of Arizona, 933 N. Cherry Tucson, AZ 85721 USA}

\author{Yi-Kuan Chiang}
\affiliation{Center for Cosmology and AstroParticle Physics (CCAPP), The Ohio State University, Columbus, OH 43210, USA}
\affiliation{Institute of Astronomy and Astrophysics, Academia Sinica, Taipei 10617, Taiwan}

\author{Roxana Popescu}
\affiliation{Department of Astronomy, University of Massachusetts, Amherst, MA 01003, USA}

\begin{abstract}
Combined observations from UV to IR wavelengths are necessary to fully account for the star-formation in galaxy clusters. Low mass $\log\mathrm{M/M_*}<10$ galaxies are typically not individualy detected, particularly at higher redshifts ($z\sim1-2$) where galaxy clusters are undergoing rapid transitions from hosting mostly active, dust-obscured star-forming galaxies to quiescent, passive galaxies. To account for these undetected galaxies, we measure the total light emerging from GALEX/$NUV$ stacks of galaxy clusters between $z=0.5-1.6$. Combined with existing measurements from \textit{Spitzer}, \textit{WISE}, and \textit{Herschel}, we study the average UV through far-infrared (IR) spectral energy distribution (SED) of clusters. From the SEDs, we measure the total stellar mass and amount of dust-obscured and unobscured star-formation arising from all cluster-member galaxies, including the low mass population. The relative fraction of unobscured star-formation we observe in the UV is consistent with what is observed in field galaxies. 
There is tentative evidence for lower than expected unobscured star-formation at $z\sim0.5$, which may arise from rapid redshift evolution in the low mass quenching efficiency in clusters reported by other studies. Finally, the GALEX data places strong constraints on derived stellar-to-halo mass ratios at $z<1$ which anti-correlate with the total halo mass, consistent with trends found from local X-ray observations of clusters. The data exhibit steeper slopes than implementations of the cluster star-formation efficiency in semi-analytical models. 
\end{abstract}

\section{Introduction} \label{sec:intro}
The secular properties of galaxies are a function of their local extra-galactic environment. Indeed, galaxies residing in over-dense clusters are distinct from the lower density ``field'' population. Relative to the field, $z<1$ clusters tend to host more massive, redder, and low star-formation rate galaxies (e.g., \citealt{Gomez2003,Baldry2006,Peng2010}), evidence for more rapid evolution at earlier times than field galaxies. Therefore, environment plays a fundamental role in setting the evolution of galaxies. Locally, ultraviolet (UV) and optical/near-infrared (IR) observations have shown that clusters have efficiently quenched their star-formation, particularly in low mass ($\mathrm{\log M/M_*}<10$) galaxies, via hot halo starvation and/or ram-pressure stripping (e.g., \citealt{guvics1,guvics4,Haines2011}). At high redshifts ($z\sim1-2$), mid- and far-IR observations have revealed a significant amount of dust-obscured star-formation occurring within galaxy clusters (e.g., \citealt{Brodwin2013, Alberts2016}).
Combined multi-wavelength studies of galaxy clusters have identified $z\sim1-2$ as a key epoch in which cluster environments transition from the sites of intense star-formation and supermassive black hole assembly to those dominated by quenched galaxies \citep{Martini2013,Alberts2014,Nantais2017}.

A stellar mass-complete and multi-wavelength census of galaxy growth in clusters is necessary to understand the important role played by environment in shaping galaxy evolution. Optical and near-IR surveys can reach stellar mass limits of $\sim10^9\,\mathrm{M_\odot}$ \citep{vanderBurg2018}; however, such sensitivity is not yet possible at UV and IR wavelengths where significant star-formation is detected in massive cluster galaxies \citep{Haines2011,Alberts2016}. To push the study of cluster members to higher redshifts while accounting for the light from less massive and faint galaxies that typically go undetected, \citealt{Alberts2021} (hereafter \alberts) developed a ``total light'' stacking method whereby near- through far-IR maps of many galaxy clusters were averaged to study the total cluster spectral energy distribution (SED) from $z\sim0.5-1.6$. They find the SED to be well-fit by a star-forming galaxy template, albeit with less warm dust than found in massive star-forming galaxies \citep{Kirkpatrick2015}, and finding that the star-formation in the average cluster from $z\sim0.5-1.6$ is predominantly occurring within $M_*/M_\odot<10^{10}$ galaxies. From observations of field galaxies $\sim30-70\%$ of the total star-formation is obscured by dust at such stellar masses \citep{Whitaker2017}; by $M_*/M_\odot\sim10^{11}$ nearly $100\%$ of the star-formation is dust-obscured. If such lower-mass galaxies contain similar amounts of dust as their field counterparts, and exhibit comparable ratios of obscured to unobscured star-formation (e.g., \citealt{Whitaker2017}), then a significant amount of star-formation could be missed by the IR observations alone. 

To account for all of the star-formation, we extend the \alberts\ methodology into the UV using GALEX coverage of the same sample drawn from the IRAC Shallow Cluster Survey. We combine the subsequent UV through IR observations to measure the average amount of dust-obscured and unobscured star-formation in clusters relative to the field. At $z<1$, the GALEX data traces the rest-frame emission from young, massive stars and provides good constraint on the mass-to-light ratio. This increases the constraint on total cluster stellar masses, the uncertainties on which were previously too large to distinguish between various star-formation efficiency scenarios with halo mass. 

This paper is organized as follows: in Section \ref{sec:data} we describe the GALEX/$NUV$ data. Section \ref{sec:methods} summarizes our total light stacking method \citep{Alberts2021}, and notable modifications for the treatment of GALEX data. Section \ref{sec:results} explores the results of combined UV through IR stacked cluster spectral energy distributions and radial profiles, which we discuss in further detail in Section \ref{sec:discussion}. Section \ref{sec:conclusion} summarizes our work, and our most notable conclusions. In this paper we adopt a $\Lambda$CDM cosmology with $h=0.7$, $\Omega_m=0.3$, and $\Lambda=0.7$.  

\section{Data} \label{sec:data}
\subsection{Cluster Sample} \label{subsec:sample}

As in \alberts, we make use of clusters identified in the B\"ootes field ($\alpha,\,\delta$=14:32:05.7,+34:16:47.5 J2000) from the IRAC Shallow Cluster Survey \citep[ISCS,][]{Eisenhardt2008}. This cluster catalog comprises $\sim$300 cluster candidates between $0.1 < z < 2$, with $\sim$100 candidates at $z>1$. Clusters are identified as overdensities of 4.5 $\mu$m flux selected galaxies in three-dimensional (RA, DEC, photometric redshift) space using a wavelet detection algorithm \citep{Brodwin2013}. At high redshift, more than $20$ clusters from the sample have been spectroscopically confirmed \citep{Stanford2005,Eisenhardt2008,Brodwin2006,Brodwin2011,Brodwin2013,Zeimann2013}, 
and the rate of contamination by spurious line-of-sight associations is expected to be $\sim10\%$ over the sample \citep{Eisenhardt2008}. From clustering measurements and halo mass ranking simulations, the median halo mass of the cluster sample is $\log \mathrm{M_{200}/M_{\odot}}\sim13.8 - 13.9$ and does not change within the redshift range of our study \citep{Brodwin2007,Lin2013, Alberts2014}. The corresponding Virial radius is R$_{200}$ $\sim$ 1 Mpc and this value is used throughout the remainder of this paper. We restrict our analysis to the redshift range $0.5~<~z~<~1.6$, with the lower bound being chosen to ensure that the angular size does not change drastically with redshift and the upper bound being chosen due to the small number of clusters in the sample at z $>$ 1.6. There are 232 clusters in the ISCS catalog within this redshift range. We make use of the photometric redshift catalogs presented in \citet{Alberts2016} for these clusters in our analysis. 

\subsection{GALEX Imaging} \label{subsec:GALEX}
The Galaxy Evolution Explorer (GALEX) satellite operated between 2003 and 2013, obtaining data in near-UV ($NUV$, $1750 - 2750$ \AA) and far-UV ($FUV$, $1350 - 1750$ \AA) wavelengths. We restrict our analysis to the $NUV$ data because (1) the $FUV$ coverage of the B\"ootes field is not uniform, with exposure times only $\sim$ 0.005 -- 0.05 times that of the $NUV$ pointings, and (2) the $FUV$ band samples largely blueward of the Lyman limit over the redshift range considered. 
Thus, the usefulness of the $FUV$ data on $z>0.5$ cluster galaxies is limited. The B\"ootes field was imaged in the $NUV$ band as part of the GALEX Deep Imaging Survey \citep[DIS,][]{Morrissey2007,Bianchi2009} in 11 pointings, each consisting of a circular exposure with diameter 1.25$^{\circ}$. Ten of eleven tiles have exposure times between $\sim20-40$ ks, whereas the tile at the center of the B\"ootes field (tile NGPDWS\_00) has an exposure time of 145 ks. The pixel scale of the tiles is 1.5\arcsec\ per pixel. The tiles overlap from $\sim$ 0.1$^\circ$--0.3$^\circ$. 


\section{`Total light' stacking of GALEX data} \label{sec:methods}
In \citetalias{Alberts2021}, we carried out total light cluster stacking of \textit{Spitzer} IRAC, \textit{WISE}, \textit{Spitzer} MIPS and \textit{Herschel} SPIRE data to construct the average infrared SED for the ISCS clusters. In the present work we add to this analysis by carrying out total cluster stacking of GALEX $NUV$ data. In this section, we briefly summarize the technique and detail the modifications made to appropriately stack GALEX data. We also describe the procedure to extract the total $NUV$ flux and radial profile measurements from the stacked images. For a detailed description of both procedures we refer the reader to \citetalias{Alberts2021}.

We process the GALEX images in a similar manner to that described in \citetalias{Alberts2021} for the \textit{WISE} $W1$/$W2$ and \textit{Spitzer} IRAC 3.6/4.5 $\mu$m images. This is because the measurement of both the total NIR and total $NUV$ fluxes from clusters suffer from similar difficulties due to individual, bright foreground sources and variable sky counts. 

We obtained the raw GALEX tiles covering the B\"ootes field from the Space Telescope Science Institute website\footnote{\url{http://galex.stsci.edu/GR6/}}. The procedure to construct the 'total light' stacks from these tiles consists of four steps: 
\begin{enumerate}
    \item Removal of bright contaminating sources (i.e. individually detectable non-cluster sources) from the individual tiles
    \item Subtract sky background from each tile (such that local sky is zero)
    \item Creation of equally sized cutouts centered on each cluster to be stacked from the processed images
    \item Stacking of the cutouts by taking the pixel-wise mean/median/weighted mean
\end{enumerate}
In order to carry out the first step, we run SExtractor \citep{Bertin1996} on the individual GALEX tiles and output the segmentation map, which marks all pixels belonging to a source\footnote{We run SExtractor with \textsc{DETECT\_THRESH} of 3.0 and \textsc{DETECT\_MINAREA} of 3; however, by varying the detection settings we have previously found that the stacked flux measurements are insensitive to them (\alberts)}.
We use the segmentation maps to mask pixels associated with detected sources from the calculations of pixel-wise 
mean/median/weighted means used to produce the final stack.

In addition to the segmentation map, we also make use of the SExtractor output catalog, which lists the sky coordinates of the detected sources. Before performing the masking of these sources, we compare this output catalog with the photometric redshift catalogs for the ISCS clusters, which gives the positions of the cluster memebers. We remove the cluster sources from the segmentation map to ensure that they remain unmasked.

To remove the variable sky background, we subtract GALEX-provided sky background maps. These background maps are created using an algorithm which is optimized for the low sky counts of UV data \citep{Morrissey2007}. The sky background in these maps is calculated over scales of 192\arcsec\ \citep{Morrissey2007}, which at the lowest redshift we consider, $z=0.5$, corresponds to 1.2 Mpc.

After processing the tiles, we create equally sized 
cutouts for each cluster, 
$27\farcm5$ to a side. This corresponds to $10.2$ ($14.2$) Mpc at $z = 0.5$ ($1.6$), which is enough to cover the entire cluster as well as the background for robust statistics. 
To stack the cutouts, we divide the clusters into four pre-defined redshift bins as detailed in \citetalias{Alberts2021}: $z$= 0.5--0.7, 0.7--1.0, 1.0--1.3 and 1.3--1.6. The bin centers and widths balance the number of clusters in each redshift window, and ensure a minimal change in angular scale per bin. Specifically, each bin contains $40-70$ clusters, each containing $\sim10-30$ cluster members with photometric redshifts \citepalias{Alberts2021}. The mean redshifts of the clusters in each bin are $z = 0.57$, 0.86, 1.13 and 1.44 respectively. The cutouts within a given redshift bin are compiled into a three-dimensional datacube over which we compute a pixel-wise weighted mean to produce the final stacked image. We do not expect to see much (if any) UV emission in the stack of the highest redshift bin, given that the GALEX $NUV$ filter lies mostly blueward of the Lyman limit at these redshifts. 

The main difference between our procedure and that of \alberts\ is the use of a weighted mean to stack the cutouts. As described in Section \ref{subsec:GALEX}, the GALEX tiles covering the B\"ootes field vary considerably in exposure times, unlike the WISE and IRAC images. Therefore, we weight each pixel by the inverse variance of the sky noise (1/$\sigma_{sky}^2$, where $\sigma_{sky}$ is the standard deviation of the sigma-clipped pixel distribution for the cutout), such that all the pixels in a given cutout have the same weight. Masked pixels do not contribute to the weighted mean or to the calculation of the weight for a cutout. 

To verify the robustness of our procedure, we also create `offset' stacks centered on random positions instead of the location of clusters. We create a random cutout associated with each cluster in a given stack, centered on a point 0.2$^\circ$ -- 0.4$^\circ$ away from the nominal cluster position, and carry out the stacking in an identical manner as for the fiducial stacks. For simplicity, the random cutout is made from the same processed GALEX image as the cluster cutout. The magnitude of the offset from the cluster center is chosen to be large enough to ensure that the offset cutout does not include any cluster signal, but not so large that the offset cutout falls outside the processed GALEX image.  None of the offset stacks exhibit detectable signal above the sky background level, suggesting that any signal in the fiducial stacks does indeed originate from the clusters and is not the result of random noise fluctuations.   

\begin{deluxetable*}{cccccccccccc}
\caption{Total cluster light photometry used in SED fits \label{tab:phot}}
\tabletypesize{\scriptsize}
\tablehead{\colhead{Redshift}&
\colhead{NUV}&
\colhead{W1}&
\colhead{W2}&
\colhead{I1}&
\colhead{I2}&
\colhead{I3}&
\colhead{I4}&
\colhead{MIPS 70}&
\colhead{S250}&
\colhead{S350}&
\colhead{S500}\\[-1.5ex]
\colhead{}&
\colhead{[$\mu$Jy]}&
\colhead{[mJy]}&
\colhead{[mJy]}&
\colhead{[mJy]}&
\colhead{[mJy]}&
\colhead{[mJy]}&
\colhead{[mJy]}&
\colhead{[mJy]}&
\colhead{[mJy]}&
\colhead{[mJy]}&
\colhead{[mJy]}
}
\startdata
\quad R=0.5 Mpc\\
$z\sim0.57$ & $7.5\pm1.3$ & $1.13\pm0.03$ & $0.60\pm0.03$ & $1.23\pm0.03$ & $0.87\pm0.03$ & $0.74\pm0.06$ & $0.82\pm0.07$ & $7.2\pm1.2$ & $56\pm12$& $33\pm10$&$16.2\pm5.9$\\[1ex]
$z\sim0.86$ & $5.9\pm0.7$ & $0.66\pm0.02$ & $0.43\pm0.02$ & $0.79\pm0.02$ & $0.55\pm0.02$ & $0.46\pm0.03$ & $0.37\pm0.03$ & $7.5\pm0.9$ & $52\pm8$&$39\pm7$&$19.8\pm3.8$\\[1ex]
$z\sim1.1$ & $2.5\pm0.9$ & $0.44\pm0.02$ & $0.37\pm0.02$ & $0.51\pm0.02$ & $0.43\pm0.02$ & $0.41\pm0.06$ & $0.35\pm0.05$ & $10.0\pm1.0$ &$66\pm11$&$58\pm8$&$29.1\pm5.2$\\[1ex]
$z\sim1.4$ & $<0.38$     & $0.19\pm0.02$ & $0.25\pm0.02$ & $0.18\pm0.02$ & $0.23\pm0.02$ & $0.20\pm0.04$ & $0.21\pm0.03$ & $3.9\pm1.1$ &$53\pm12$&$50\pm9$&$30.4\pm6.2$ \\[1ex]
\hline 
\quad R=1.0 Mpc\\
$z\sim0.57$ & $9.8\pm3.2$& $1.47\pm0.10$ & $0.74\pm0.09$ & $1.43\pm0.10$ & $1.05\pm0.11$ & $0.94\pm0.21$ & $1.0\pm0.28$  & $9.2\pm4.0$ & $74\pm29$&$43\pm21$&$21.8\pm13.3$ \\[1ex]
$z\sim0.86$ & $8.6\pm2.6$& $0.73\pm0.05$ & $0.49\pm0.05$ & $0.92\pm0.07$ & $0.62\pm0.08$ & $0.48\pm0.13$ & $0.42\pm0.11$ & $8.7\pm2.7$ & $76\pm16$  &$57\pm14$&$28.9\pm9.2$\\[1ex]
$z\sim1.1$ & $<1$ & $0.53\pm0.05$       & $0.45\pm0.05$ & $0.65\pm0.06$ & $0.53\pm0.07$ & $0.53\pm0.23$ & $0.33\pm0.16$ & $9.1\pm2.7$ & $92\pm22$ &$84\pm17$&$42.1\pm9.9$\\[1ex]
$z\sim1.4$ & $<0.8$  & $0.17\pm0.06$    & $0.24\pm0.08$ & $0.13\pm0.06$ & $0.21\pm0.07$ & $0.20\pm0.15$ & $0.23\pm0.10$ & $6.6\pm3.2$ & $70\pm28$ &$70\pm21$&$44.4\pm11.8$\\[1ex]
\enddata
\end{deluxetable*}

\subsection{Measurement of UV flux and radial profiles} \label{subsec:radial profile}

From each map, we perform aperture photometry using a series of circular apertures to measure the cumulative flux profile as a function of cluster-centric radius. In addition, we measure the flux in progressively larger annuli to find the radial surface brightness profiles. 

We place the center of the apertures at the image center, which corresponds to the nominal cluster center. In practice, there is uncertainty in the cluster centroiding on the order of $\sim$ 15\arcsec\ (10 pixels) from the 15\arcsec\ pixel scale of the density maps used to select the ISCS clusters. \citet{Gonzalez2019} carry out a detailed analysis for the MaDCoWS cluster sample, which is also selected from density maps with a pixel scale of 15\arcsec\, and confirm that the scatter in cluster position is $\approx$ 15\arcsec. Given that this uncertainity is small, we do not expect it to significantly affect the cumulative flux; however, there could be some effect on the differential flux profile. 
We investigate the effects of centroiding uncertainty in \alberts, and find that the overall effect on the observed radial profile as compared to the intrinsic radial profile is a depression in the innermost radial bins and an excess at 20\arcsec\ -- 40$\arcsec$. 

We measure and subtract out sky signal by taking the outlier-resistant mean of the pixel values in an annular bin between $150^{\prime\prime}-200$\arcsec\ in radius.
We then measure the aperture flux and surface brightness in circular and annular apertures respectively, starting at a radius of 5.5\arcsec\ and with an increment of 5.5\arcsec. We choose this step size in order to directly compare our results with the NIR radial profiles of the ISCS clusters from \citetalias{Alberts2021}. 
The cumulative flux for a given radial bin is the sum of all pixel values within the aperture, while the surface brightness is the mean of all pixels within the annulus. As such, the cumulative flux is measured in units of $\mu$Jy, while the differential flux is measured in units of $\mu$Jy/pix.

We calculate the sigma-clipped standard deviation of the pixel distribution, again within an annulus of radius 150\arcsec\--200\arcsec\, and assume this to be the per-pixel uncertainty $\sigma_{pix}$. In addition, we calculate the uncertainty in the measured sky level $\sigma_{sky}$ by placing 500 annular apertures at random points in the offset stack and calculating the standard deviation of the resulting distribution of mean sky values. To each measurement of the cumulative flux, made in a circular aperture containing $N_{circ}$ pixels, we assign an error of $\sigma_{\rm f}^2 = N_{circ}\sigma_{pix}^2 + N_{circ}^2\sigma_{sky}^2$. To each measurement of the surface brightness, made in an annulus containing $N_{annulus}$ pixels, we assign an error of  $\sigma_{\rm SB}^2 = (\sigma_{pix}^2+\sigma_{rms}^2)/N_{annulus} + \sigma_{sky}^2$, where $\sigma_{rms}$ is the standard deviation of the pixel values in the annular aperture. 

In the two lowest redshift bins, we detect signal in the aperture-integrated GALEX/$NUV$ photometry at $\gtrsim5\sigma$ within $R=0.5$ Mpc, and at a similar significance within $R=1$ Mpc, approximately the Virial radius. At $z\sim1.1$, we report a marginal detection of the cluster-associated UV light at $2.5\sigma$ within $R=0.5$ Mpc. We do not detect any signal in the $NUV$ stack in the highest redshift bin, as expected, given that at $z>1.3$ the majority of the GALEX/$NUV$ band falls blueward of the Lyman limit. In the third redshift bin, where the detection is marginal, we place a $3\sigma$ upper limit on the UV flux using $3\sigma_{\mathrm{sky}}$. 
Aperture photometry is listed in Table \ref{tab:phot}, including fluxes measured within 0.5 and 1 Mpc for the GALEX data and for the longer wavelengths stacks from \citetalias{Alberts2021}.

\section{Results\label{sec:results}}

\subsection{Total Cluster Flux and Spectral Energy Distributions}
\begin{figure*}
    \centering
    \includegraphics[width=\textwidth]{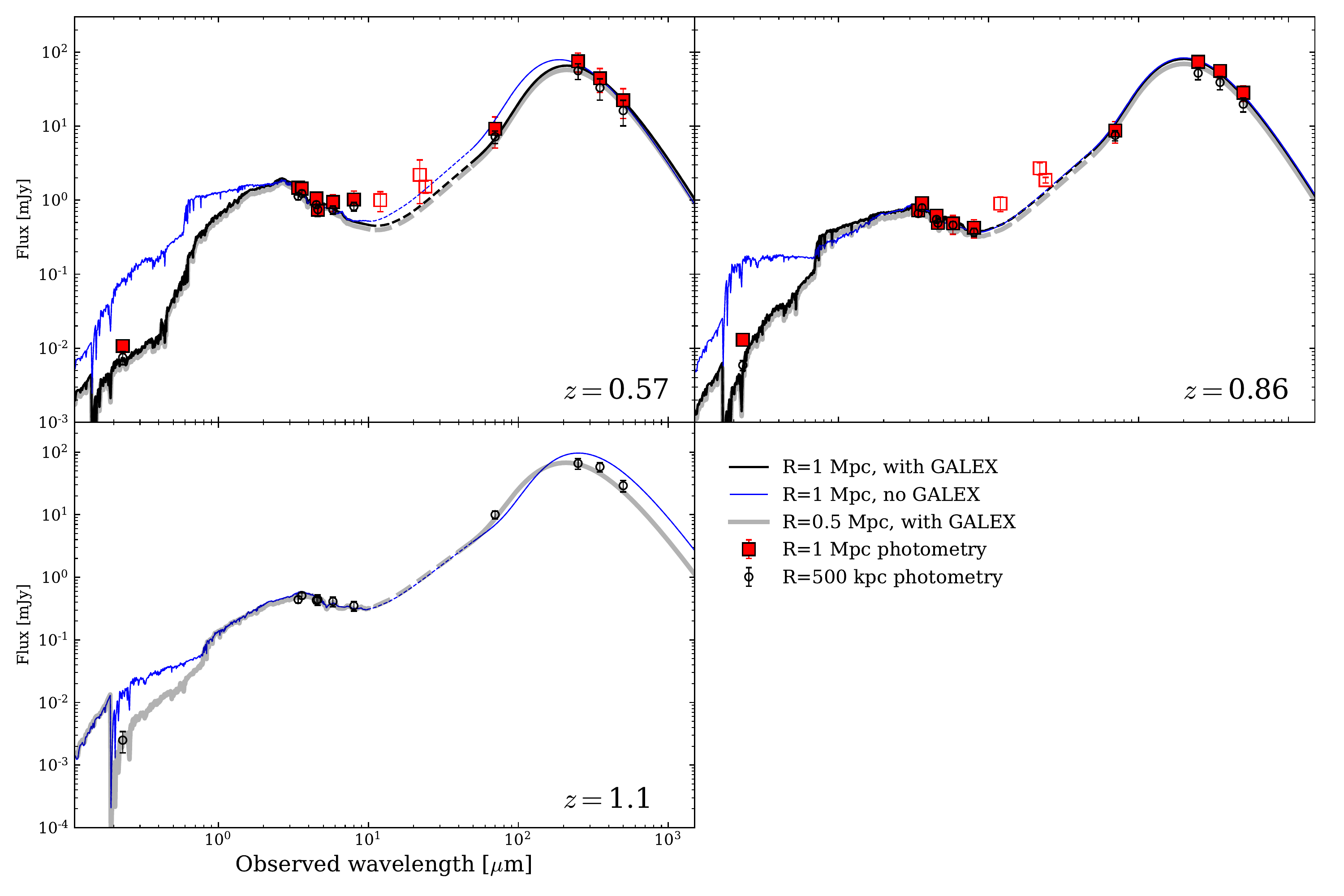}
    \vspace{-10pt}
    \caption{SED fits from \texttt{CIGALE} to the total cluster light in the three redshift stacks for which we report GALEX detections. 
    Note that we do not include stacked \textit{Spitzer}/$MIPS$ 24\,\micron\ emission and $W3+W4$ emission (open red squares) in our SED fitting as we do not fit for PAH emission because the redshift bins are too broad to capture the PAH shape. The SEDs in this wavelength regime are dashed for clarity. Photometry included in the SED fits are shown as solid red squares (R$=1$ Mpc apertures) and as open black cirlces (R$=0.5$ Mpc apertures).
    We show the added constraint from the GALEX $NUV$ photometry on the shape of the SED by comparing to the \texttt{CIGALE} results from \alberts\ (blue), which fit all but the UV data. Omitting UV photometry from the fit over-predicts the UV emission. 
    }
    \label{fig:seds}
\end{figure*}

Using the aperture-integrated photometry, we fit the UV-to-IR SEDs of the average cluster stack at each redshift using \texttt{CIGALE} \citep{Burgarella2005,Noll2009,Boquien2019}, a stellar population synthesis code that assumes energy conservation in dust-reprocessing to also model the far-IR SED. In this work, we update the SED fits from \alberts\ where we have UV detections to further constrain stellar masses and the unobscured star-formation rates. Specifically, we fit both the R$=0.5$ Mpc and R$=1$ Mpc aperture photometry at $z\sim0.57$ and $z\sim0.86$. For the $z\sim1.1$ bin, we only fit the R$=0.5$ Mpc data where we report a $\sim2.5\sigma$ detection of UV signal (Table \ref{tab:phot}). The SED fit parameters are mostly the same as described in \alberts: we model the dust emission following \cite{Casey2012} as a single-temperature greybody plus a power-law spanning the mid-IR regime. We omit the $W3$, $W4$, and {\it MIPS} $24\,\mu$m stacked photometry from the SED fits as they probe the broad Polycyclic Aromatic Hydrocarbon (PAH) features which, owing to our large redshift bin widths of $\Delta z=0.3$,  are diluted in the stacked spectrum leading to a $\sim10\%$ systematic uncertainty on the broadband photometry (see Section~4.2 in \alberts). We assume a fixed dust-emissivity index of $\beta=1.5$, a mid-IR power-law slope of $\alpha=2$, and a star-formation history with short ($1-5$ Myr) and long ($100-1000$ Myr) decaying $e$-folding factors to model an initial period of star-formation followed by a  burst with an onset at later times. The boundaries on star-formation history priors are expanded relative to the fits described in \alberts\ because of the added UV photometry's sensitivity to, among other quantities, the age of the young starburst population; however, all fits favor zero contribution from the short-burst mode (i.e., short-burst mass fractions converge to zero). The best-fit SEDs for each redshift bin are shown in Figure \ref{fig:seds}. Derived properties and their corresponding uncertainties are given in Table \ref{tab:seds}.

\subsubsection{Derived stellar mass estimates with $NUV$ data}
In \alberts, we presented stellar mass estimates derived from the $R=1$ Mpc photometry. Here we update these estimates with the new GALEX/$NUV$ values using aperture-integrated photometry at both $R=0.5$ Mpc and $R=1$ Mpc. We expect to improve the uncertainties on \mstar\ as the $NUV$ data constrains recent star-formation that can influence the stellar mass to light ratio; indeed, we report improvements at $z<1$ where the UV signal is detected at $>3\sigma$. 
Specifically, 
the $\mathrm{M_*}$ uncertainties at $z\sim0.57$ and $z\sim0.86$ improved by $\sim25-40\%$ when the UV data is included in the SED fitting of 1 Mpc aperture photometry (Table \ref{tab:seds}). 
In general, stellar masses measured with and without the $NUV$ data are consistent within $1\sigma$. Similarly, stellar masses within 0.5 Mpc and 1 Mpc are in good agreement at $z<1$, suggesting that the 0.5 Mpc aperture traces most of the stellar mass at $z\sim1.1$.

\begin{deluxetable*}{lccccccccc}
\tablecolumns{10}
\tabletypesize{\scriptsize}
\tablecaption{Derived Properties from SED Fitting \label{tab:seds}}
\tablehead{\colhead{$z$} & \colhead{R$^{a}$} & \colhead{$\mathrm{M_{*}}$ (with UV)}&
\colhead{$\mathrm{M_{*}}$ (no UV)}& 
\colhead{$\mathrm{L_{IR}}$} & 
\colhead{SFR$_{\mathrm{UV}}$} &
\colhead{SFR$_{\mathrm{IR}}$} &
\colhead{SFR$_{\mathrm{Tot}}$} & 
\colhead{$f_{obs}$} &
\colhead{sSFR} \\
\colhead{} &
\colhead{[Mpc]} & 
\colhead{[$\mathrm{10^{11\,}M_\odot}$]} & 
\colhead{[$\mathrm{10^{11\,}M_\odot}$]} & 
\colhead{[$\mathrm{10^{11}\,L_\odot}$]}&
\colhead{[$\mathrm{M_{\odot}\, yr^{-1}}$]}&
\colhead{[$\mathrm{M_{\odot}\, yr^{-1}}$]}&
\colhead{[$\mathrm{M_{\odot}\, yr^{-1}}$]}&
\colhead{ }&
\colhead{[$\mathrm{10^{-10}\, yr^{-1}}$]}
} 
\startdata 
$z\sim0.57$ & $0.5$ & $13.9\pm7.0$ & \nodata & $4.9\pm2.4$ & $7\pm2$ & $73\pm17$ & $80\pm17$  & $0.91\pm0.20$ & $0.6\pm0.2$\\[1ex]
 			& $1.0$ & $13.9\pm4.5$ & $14\pm6$& $6.4\pm1.5$ & $7\pm4$ & $96\pm23$ & $103\pm23$ & $0.93\pm0.31$ & $0.7\pm0.3$\\[1ex]
\hline 
$z\sim0.86$ & $0.5$ & $7.4\pm2.0$ & \nodata & $15.2\pm3.0$ & $36\pm6$  & $228\pm53$ & $264\pm55$ & $0.86\pm0.26$& $3.6\pm1.2$\\[1ex]
			& $1.0$ & $7.3\pm2.3$ & $9\pm6$ & $18.2\pm3.3$ & $42\pm10$ & $273\pm50$ & $315\pm51$ & $0.87\pm0.21$& $3.7\pm1.4$\\[1ex]
\hline 
$z\sim1.1$	& $0.5$ & $14.3\pm12.5$ & \nodata 	& $32.3\pm3.2$ & $35\pm8$ 	& $499\pm57$ & $534\pm58$ & $0.93\pm0.16$  & $3.7\pm3.6$\\[1ex]
			& $1.0$ & \nodata 		& $15\pm15$ & $37.8\pm6.2$ & $<120$ 	& $567\pm92$ & $687\pm96$ & $>0.80\pm0.18$ & \nodata\\[1ex]
\hline 
$z\sim1.4$	& $0.5$ & \nodata & \nodata & $36.5\pm4.9$ & $<10$ & $536\pm75$  & $557\pm75$  & $>0.98\pm0.20$  & \nodata\\[1ex]
			& $1.0$ & \nodata & $7\pm8$ & $43.7\pm9.2$ & $<15$ & $665\pm138$ & $680\pm138$ &  $>0.97\pm0.28$ & \nodata \\[1ex]
\enddata
\tablenotetext{a}{ Aperture radius used in measuring integrated photometry in the stack maps.}
\end{deluxetable*}

\subsection{The Fraction of Dust-obscured Star-formation in Clusters}
From the UV and IR stacked photometry and fits to the cluster SED we measure dust-obscured and unobscured star-formation rates 
to test the evolution in dust obscuration in clusters between $z\sim0.5-1.1$. We infer a dust-obscured SFR (\sfrIR) from the IR luminosity output by CIGALE, which is well-constrained by the \textit{Herschel} photometry, using \sfrIR$/[\mathrm{M_\odot\,yr^{-1}}]=1.5\times10^{-10}\,\mathrm{L_{IR}/L_\odot}$ \citep{Murphy2011}. The values we report on \lir\ and \sfrIR\ are consistent with \alberts\ within $1\sigma$. From the UV data, we measure an unobscured SFR (\sfrUV) by first inferring a rest-frame GALEX/$FUV$ luminosity from the best-fit SED which we then convert into a star-formation rate using \sfrUV$/[\mathrm{M_\odot\,yr^{-1}}]=4.42\times10^{-44}\,\mathrm{L_{FUV}/[erg\,s^{-1}]}$ \citep{Murphy2011}. The SFR calibrations used to measure \sfrIR\ and \sfrUV\ are both derived from calibrations against extinction-free 33 GHz measurements of the SFR \citep{Murphy2011}, and are therefore aptly suited for empirically estimating total star-formation rates in dusty systems like the cluster-memebers in our study (\alberts). While CIGALE returns time-averaged star-formation rates from the best-fit star-formation history, we use empirical calibrations when measuring \sfrUV\ and \sfrIR\ to derive independent obscured and unobscured star-formation rates. In this manner we avoid biasing our SFR measurements by the particular choice of star-formation history shape which can introduce $\sim15\%$ systematic uncertainty on measured SFRs (e.g., \citealt{Buat2014}). We note that $1\sigma$ errors on SFRs in Tab.~\ref{tab:seds} from observational and calibration uncertainties are generally $>15\%$.

From \sfrIR\ and \sfrUV\ measured using the empirical calibrations of \cite{Murphy2011}, we compute the fraction of dust-obscured star-formation within each cluster, first defined by \cite{Whitaker2017} as 
\begin{equation}
    f_{obs}\equiv \frac{\mathrm{SFR_{IR}}}{\mathrm{SFR_{IR}+SFR_{UV}}}
\end{equation}
which is reported alongside the stellar masses and IR luminosities in Table \ref{tab:seds} with propagated systematic and measurement $1\sigma$ uncertainties. The total star-formation rates measured in each redshift bin are predominantly dust-obscured ($f_{obs}>0.8$).

The sensitivity of the GALEX/$NUV$ filter to the unobscured SFR in the highest redshift bin is diminished by the sampling of wavelengths below the Lyman limit for any cluster at $z\gtrsim1.45$ input into the stack. Indeed, of the 40 clusters in this redshift bin, 13 have $z>1.45$. To test whether or not our reported $3\sigma$ upper limit is still constraining for \sfrUV, we re-fit the 1 Mpc $z\sim1.4$ SED while treating the upper limit as a data point to measure the maximum values of \sfrUV\ allowable by the observations. From this fit, we measure \sfrUV$\,=1.5M_\odot~{\rm yr}^{-1}$ which corresponds to $f_{obs}\sim1$. Thus, we are unlikely to be missing a significant amount of unobscured star-formation even in the highest redshift bins due to the rest-frame sampling of the SED by the $NUV$ filter. 

\begin{figure*}
    \centering
    \gridline{\fig{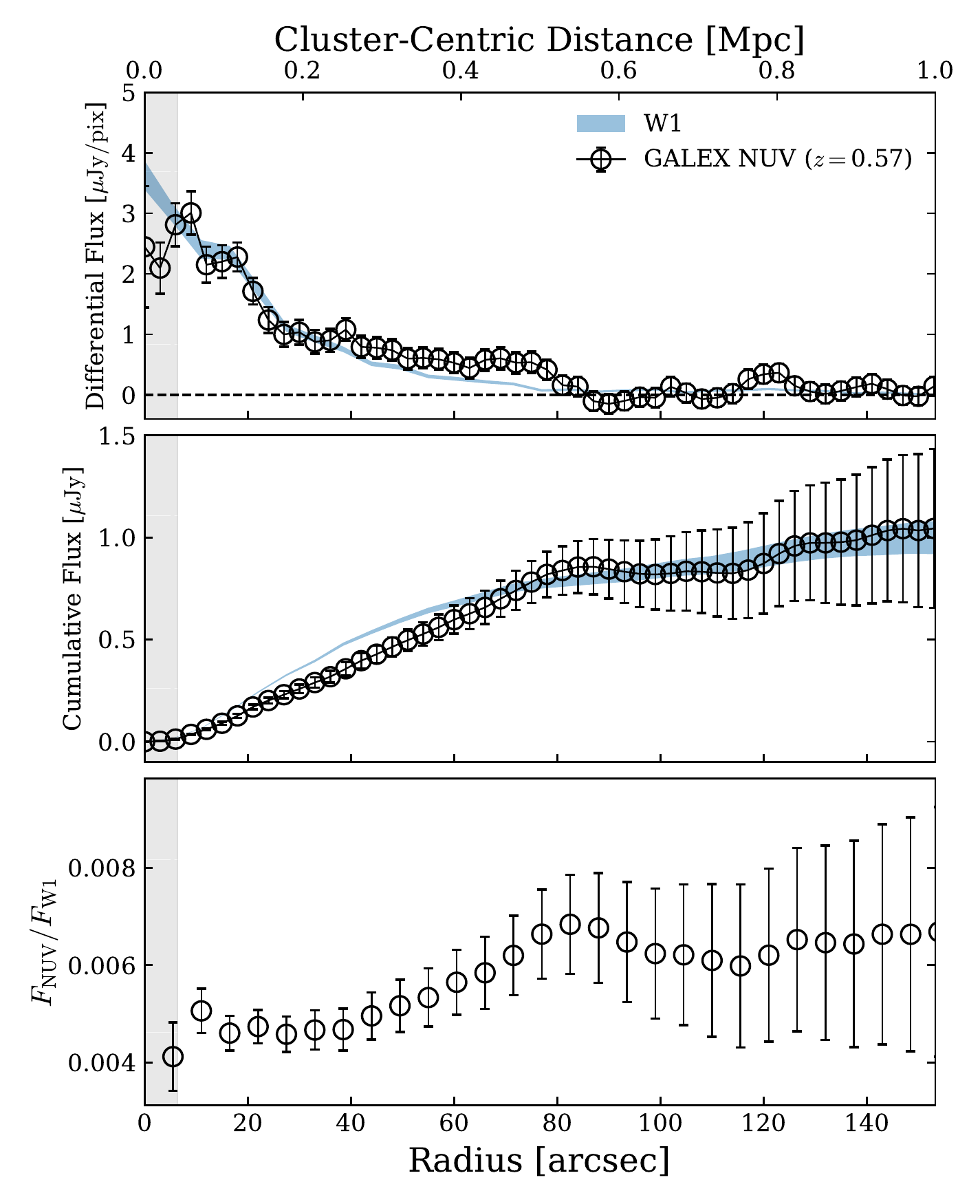}{0.49\textwidth}{}
    \fig{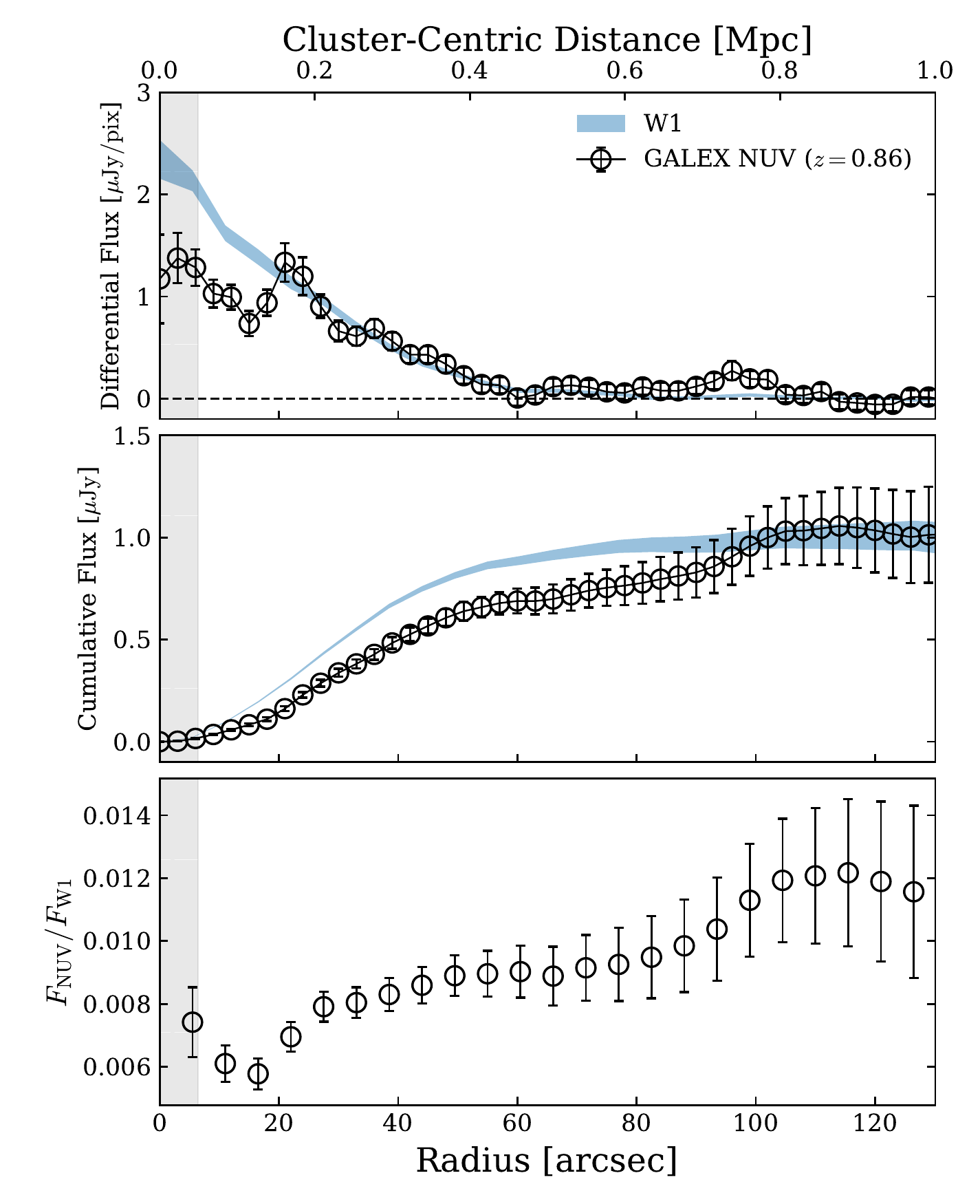}{0.49\textwidth}{}}\vspace{-30pt}   
    \caption{Radial profiles measured from the stacked GALEX/$NUV$ maps at $0.5<z<0.7$ (\textit{Left}) and $0.7<z<1.0$ (\textit{Right}). We compare the differential and cumulative UV flux radial profiles (black) against those measured from the $W1$ stacks (shaded blue) in the upper and middle panels respectively. Shaded gray regions on all panels correspond to the $6\farcs4$ FWHM of $W2$. All data in the top panels have been normalized to unity at 200 kpc. The cumulative flux profiles have been normalized at 1 Mpc (see Table \ref{tab:phot} for absolute values). The bottom panels show the ratio between the GALEX/$NUV$ and $W1$ cumulative flux profiles (no normalization). At $z\sim0.56$, the $NUV$ and $W1$ cumulative profiles are well-correlated across the extent of the cluster, whereas the differential $NUV$ profile exhibits a 3$\sigma$ discrepancy in the innermost bin relative to $W1$. At $z\sim0.86$, we observe a deficit in $NUV$ signal relative to the $W1$ profile at radii below $\sim0.2$ Mpc, reminiscent of the suppressed \sfrIR\ found in \alberts\ for all redshifts. }
    \label{fig:uv_wise}
\end{figure*}

\subsection{Radial Flux Profiles}
The radial distribution in the stacked GALEX maps traces the distribution of unobscured star-formation in clusters. \alberts\ reported evidence for suppressed dust-obscured star-formation at $R<0.3-0.5$ Mpc relative to the stellar mass profiles from $0.5<z<1.6$; we now test whether or not this also extends to the unobscured stellar light by comparing the GALEX/\textit{NUV} radial profiles with the multiwavelength profiles from \alberts. The GALEX/\textit{NUV} surface brightness measurements in the two highest redshift bins do not have sufficient S/N, and we therefore restrict our radial analysis to the two lowest redshift bins.

\subsubsection{GALEX/$NUV$ vs. $W1$}
As discussed in detail by \alberts\ (see their Section 4.1.2), comparisons between radial profile measurments made at different wavelengths are biased by differing sky noise subtraction techniques, centroiding uncertainties, and differences in each instruments' PSF. To mitigate these uncertainties, we begin by comparing GALEX/\textit{NUV} which traces the unobscured star-formation against $W1$ and $W2$, tracers of stellar mass. The FWHM of the GALEX/\textit{NUV} beam is $4\farcs9$, which is comparable to \textit{WISE} ($6\farcs1-6\farcs4$); moreover, the sky is subtracted on the same scales in all three wavelength stacks. 

Figure \ref{fig:uv_wise} shows the differential and cumulative flux (F$_{\mathrm{NUV}}$) radial profiles from the GALEX stacks in the two first redshift bins, compared against $W1$. Note that the $W2$ radial profile is nearly indistinguishable from $W1$ within $1\sigma$ errors in both cases (see Figure 4 of \alberts). At both redshifts, the $W1$ and $NUV$ radial profiles are similar at larger radii, but exhibit a deficiency in UV light in the innermost $\sim15$\arcsec$-30$\arcsec\ relative to the $W1$ profile when normalized at 1 Mpc. 
The ratio in the cumulative flux profile rises from the innermost region out to $\sim80$\arcsec, reminiscent of the evidence for suppressed $\mathrm{F_{250}/F_{W1}}\sim$\,\sfrIR$/\mathrm{M_*}$ in the cluster centers reported by \alberts\ for all of the redshift bins. This trend is more pronounced at $z\sim0.86$, which could be indicative of evolving quenching scenarios. In summary, there is evidence of suppressed obscured and unobscured star-formation in the center of our $z<1$ cluster stacks relative to the stellar mass.  

\begin{deluxetable}{lccccc}
\tabletypesize{\scriptsize}
\caption{Results from Radial Profile Fits to GALEX/$NUV$ and SPIRE 250$\,\mu$m maps. \label{tab:bic}}
\tablehead{\colhead{Band} &
\colhead{$z$} &
\colhead{$\sigma_{1G}$ [\arcsec]}& 
\colhead{$\sigma_{2G}$ [\arcsec]}& 
\colhead{$|\Delta$BIC$|$}&
\colhead{Preferred}
}
\startdata
UV$_{\mathrm{mean}}$ & 0.57 & 34 & (15,48) & 72 & Double Gaussian \\[1ex]
UV$_{\mathrm{med}}$ & 0.57 & 34 & (20,50) & 113 & Double Gaussian  \\[1ex]
SPIRE & 0.57 &  41 & (1,41) & 6.4 & Single Gaussian \\ [1ex]
\hline
UV$_{\mathrm{mean}}$ & 0.86 & 31 & (28,62) & 1.3 & Single Gaussian \\[1ex]
UV$_{\mathrm{med}}$ & 0.86 & 33 & (25,61) & 133 & Double Gaussian  \\[1ex]
SPIRE & 0.86 &  37 & (3,37) & 6 & Single Gaussian \\ [1ex]
\enddata
\end{deluxetable}

\subsubsection{GALEX/$NUV$ vs. SPIRE\label{sec:nuv_spire}}
To test for differences or similarities in the structure of dust-osbscured vs. unobscured star-formation across the clusters, we compare the $NUV$ to SPIRE $250\,\mu m$ radial profiles at $z\sim0.57$ and $z\sim0.86$. First, we convolve the $NUV$ stack maps with the SPIRE $250\,\mu m$ beam and measure both mean and median radial profiles at a FWMH$\,=18\farcs1$ spatial resolution; however, the following conclusions remain the same if we use the native $NUV$ mean-weighted radial profiles.

To quantitatively test for differences in the radial distribution between unobscured and obscured star-formation, we devise a non-parametric test of the relative contribution from a central and extended emission component to the total differential flux profile. First, we fit a single Gaussian to the SPIRE beam-convolved $NUV$ profile and the SPIRE 250$\,\mu$m profile. As shown in Table \ref{tab:bic}, these fits have $\sigma$ widths between $\sim30-40$\arcsec. Next, we add another Gaussian component to the fit, which we restrict to $\sigma<30$\arcsec\ and find the best fit for this two-component Gaussian model. By calculating Bayesian information criteria (BIC\footnote{$\mathrm{BIC=N\ln(\chi^2/N)+\ln(N)N_{var}}$ for N data points and $N_{var}$ free parameters.}), and the change in BIC between model fits ($|\Delta$BIC$|$), we can test which of a single or two-component Gaussian model is preferred accounting for the overall fit to the data while penalizing models with more free parameters. The results from each set of fits are shown in Table \ref{tab:bic}. While a given $|\Delta$BIC$|$ does not equate to a particular $p-$value in a straightforward, or easy to calculate manner, differences in BIC at the level of $|\Delta$BIC$|>6$ are considered strong, whereas $3\lesssim|\Delta$BIC$|<6$ are marginal to likely \citep{KaasRaftery1995}. Values of $|\Delta$BIC$|>10$ are considered very strong. To test whether or not these measurements are sensitive to bright central galaxies that may have been missed by the masking procedure, we repeat the BIC analysis on the median radial profiles convolved with the SPIRE beam, which is insensitive to the pixels belonging to bright central sources.

\begin{figure*}
    \gridline{\fig{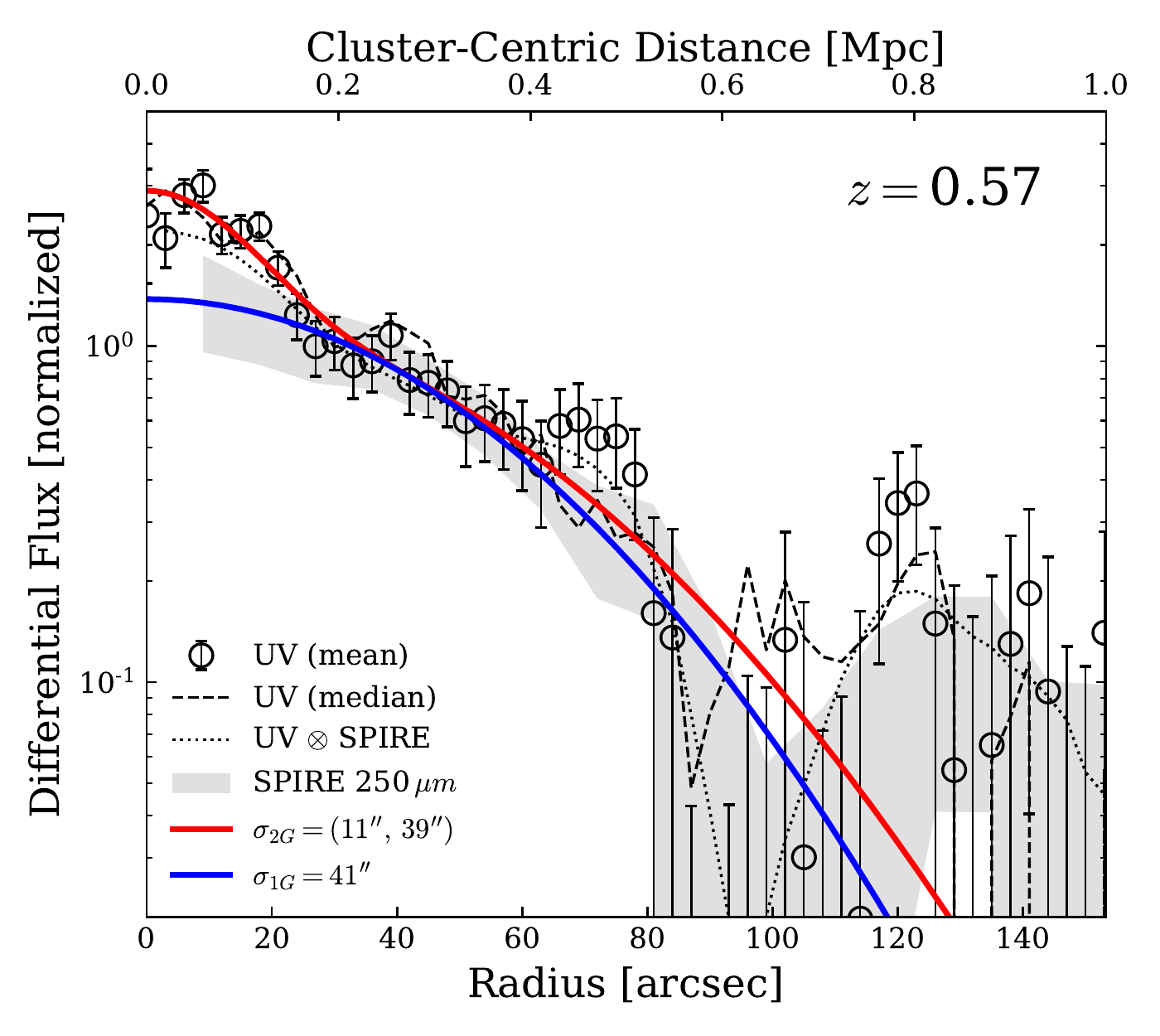}{0.45\textwidth}{}
    \fig{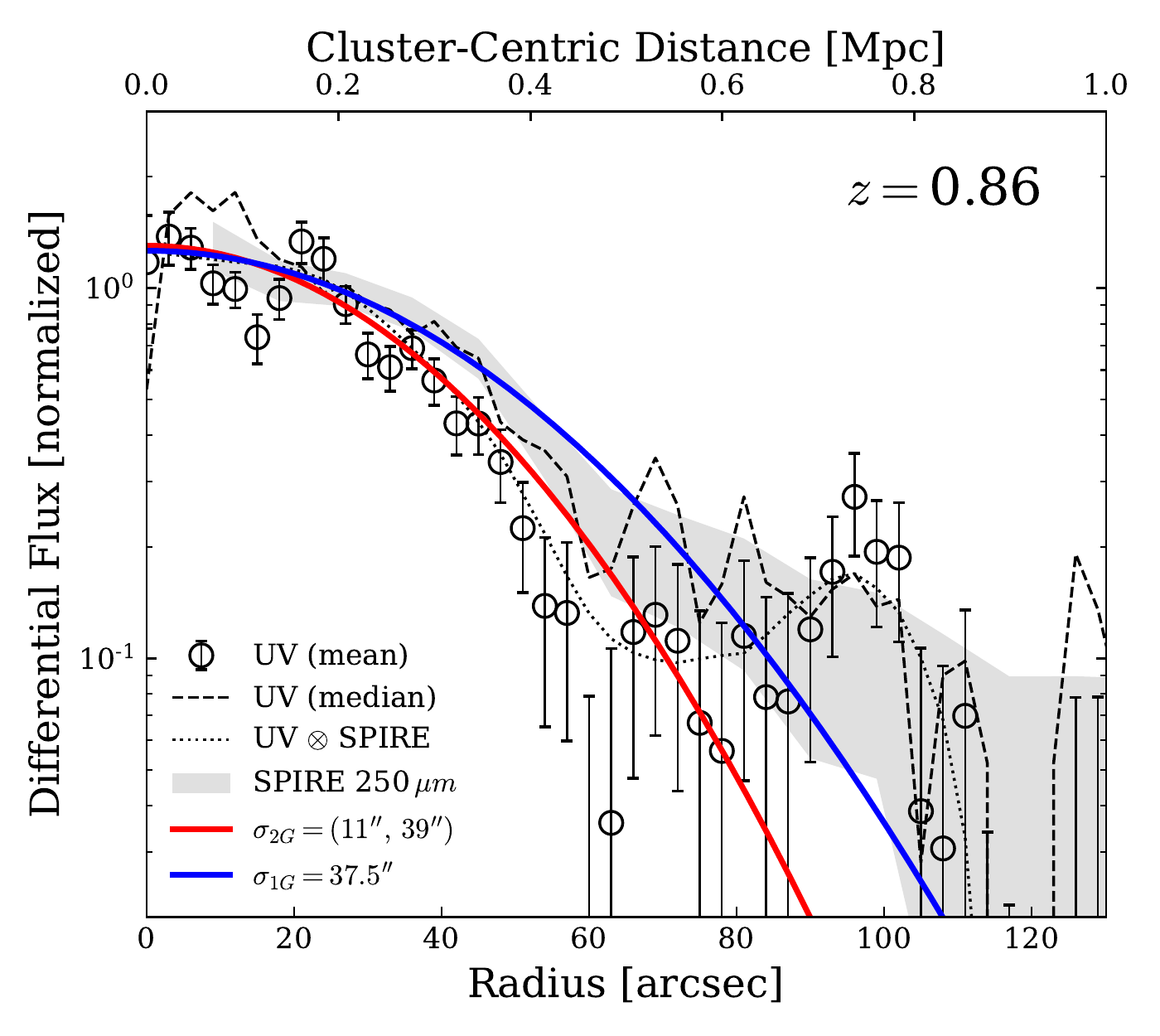}{0.45\textwidth}{}
    }\vspace{-20pt}
\caption{Differential flux radial profiles for the total cluster light in the GALEX $NUV$  filter and \textit{Herschel}/SPIRE 250\,\micron\ band (shaded gray) for the first two redshift bins with UV detections in the stacks: (\textit{Left}) $z=0.57$, (\textit{Right}) $z=0.86$. We show the results from the mean UV (open circles), and median UV (dashed) stacked maps, as well as the mean UV stacks convolved with the PSF of the SPIRE 250$\,\mu$m band (dotted). Best fits Gaussian models to the GALEX mean-convolved stacks are shown in red. Solid blue lines corresond to the best fit single Gaussian model to the SPIRE data.  Each radial profile is normalized at 0.2 Mpc. 
The UV light is more centrally concentrated than the IR emission in the $z\sim0.57$ stacks, while the evidence at $z\sim0.86$ remains elusive. }
\label{fig:radial_profiles}
\end{figure*}

Acknowledging the caveats of the BIC in model selection (see \citealt{Shi2012} for a concise review), the GALEX/$NUV$ stacks at $z\sim0.57$ exhibits very strong evidence for a two-component model, whereas a single-component model is strongly favored for the SPIRE $250\,\mu$m data (Figure \ref{fig:radial_profiles}). This suggests that the underlying distribution in unobscured star-formation across an average cluster at $z\sim0.57$ contains a component that is more centrally concentrated relative to the distribution in dust-obscured star-formation. At $z\sim0.86$, the two-component model is strongly favored for the median UV profile while the evidence for or against either model is marginal in the mean profile. As is the case for the lowest redshift bin, the SPIRE radial profile at $z\sim0.86$ prefers a single component. In addition to containing this centrally concentrated component, both $z<1$ UV profiles exhibit deficits at $<0.2$ Mpc with respect to the \textit{WISE} profiles tracing the stellar mass distribution, which is also seen for SPIRE relative to \textit{WISE} (\alberts). Thus, the data suggests an overall deficit in both star-formation-rate tracers for the inner-most regions ($<0.2-0.5$ Mpc) of $z<1$ galaxy clusters, and that the obscured star-formation is more extended than the unobscured component.

\section{Discussion}\label{sec:discussion}
In this work, we report the first complete census of star-formation in high-$z$ galaxy clusters accounting for both the obscured and unobscured components through combined UV and IR stacks. The mean halo mass of stacked clusters is $10^{13.9}\,\mathrm{M_\odot}$, corresponding to a Virial radius of $\mathrm{R_{200}\sim1\,Mpc}$. We find that the total star-formation rate per $z\sim0.5-1.4$ cluster, on average, is predominantly dust-obscured and ranges from $\sim100-600\,\mathrm{M_\odot\,yr^{-1}}$, rising steeply with redshift. The GALEX/$NUV$ measurements also place strong constraints on derived stellar masses at $z<1$. 

\subsection{Dust-obscured star-formation in low mass galaxies \label{sec:dustObsSfr}}
By comparing the Herschel total light stacks to stacks of individually identified massive ($\log\mathrm{M/M_\odot}>10.1$) cluster galaxies and analysing their near- to far-IR SEDs, \alberts\ argued that the total IR light in clusters is dominated by low mass galaxies, which in the field, have more unobscured SF on average relative to higher mass galaxies \citep{Whitaker2017}. From this conclusion, one may expect the GALEX/$NUV$ data to reveal a substantial amount of \sfrUV\ originating from low mass, low dust galaxies, if low mass cluster galaxies have similar $f_{obs}$ to field galaxies. 

To test whether or not the amount of unobscured SF we observe is consistent with \alberts\ and the $f_{obs}(M_*)$ measured in the field,  we compare measured \sfrUV\ with predicted values. We begin by assuming that the fraction of \sfrIR\ arising from the low mass galaxies is $\sim70\%$ as measured by \alberts. Next, we predict the expected \sfrUV\ if the obscured fraction in cluster galaxies is the same as in the field using the relation from \cite{Whitaker2017}\footnote{Note that \cite{Whitaker2017} reports two fits for $f_{obs}(M_*)$ made with different empirical SFR calibrations. We use the results made from the SFR conversions from \cite{Murphy2011} to match our SFR calculations made from the same relations.}. The mean obscured fraction is then:
\begin{equation}
   \left \langle\frac{1- f_{obs}}{ f_{obs}}\right\rangle =\frac{\int [1-f_{obs}(M_*)]\mathrm{SFR(M_*,z)}\psi(M_*,z) dM_*}{\int f_{obs}(M_*)\mathrm{SFR(M_*,z)}\psi(M_*,z) dM_*}
   \label{eq:fobs}
\end{equation}
where the stellar mass function, $\psi$, is taken from \cite{Marchesini2009} and the SFR-$\mathrm{M_*}$ main sequence from \cite{Speagle2014}.

\begin{figure}
    \centering
    \includegraphics[width=.44\textwidth]{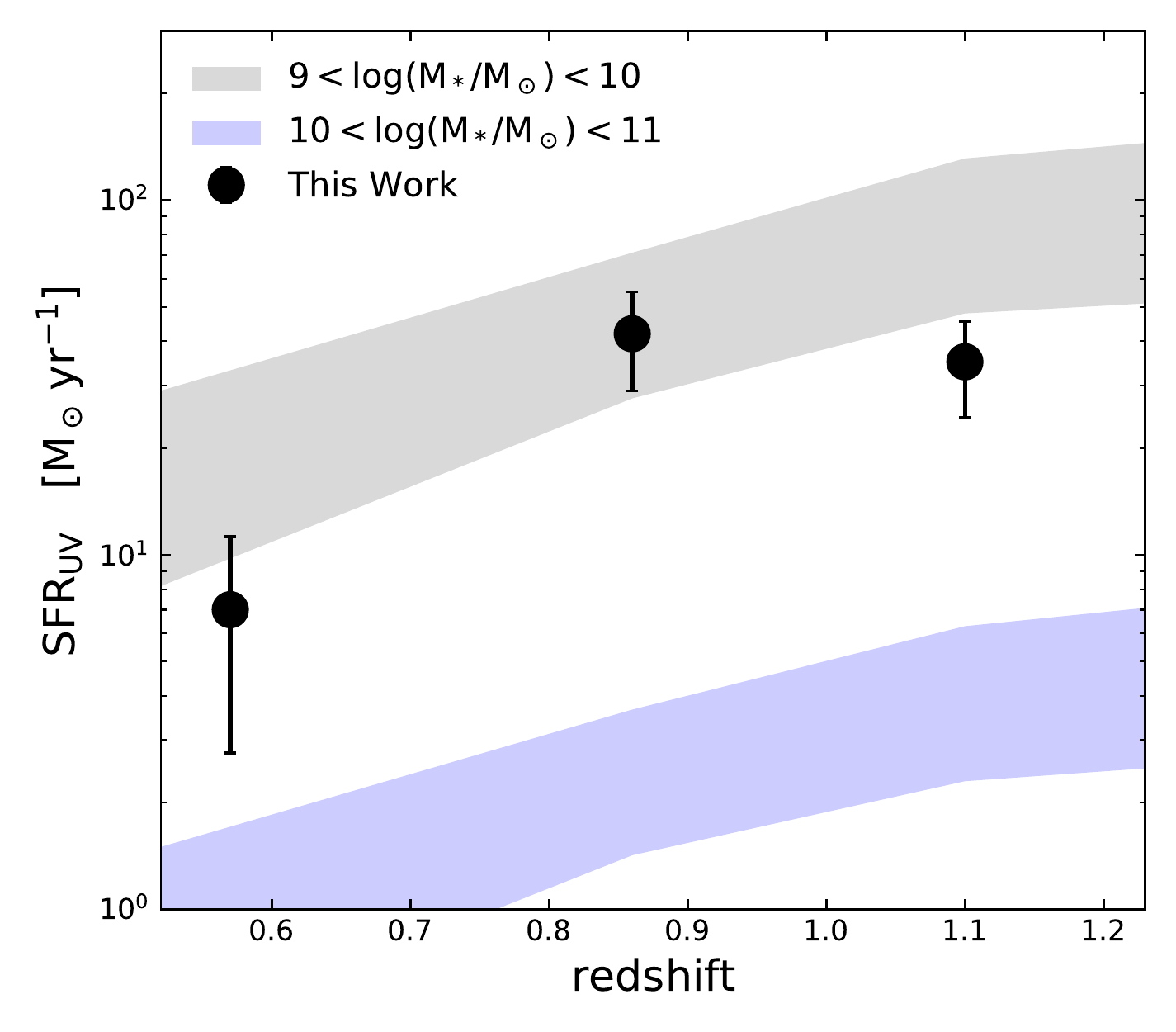}
    \caption{Unobscured star-formation rates from $z\sim0.5-1.1$ from our stacks of galaxy clusters, measured through a $R<1$ Mpc aperture (black circles). Upper limits are $3\sigma$. The shaded grey regions encapsulates the $3\sigma$ domain of predicted unobscured star-formation rates arising from $10^{9}<M_*/M_\odot<10^{10}$ galaxies assuming their fractional contribution to \sfrIR\ of 70\% \citep{Alberts2021}, and that the obscured fraction of star-formation is field-like \citep{Whitaker2017}. The blue shaded region represents the $3\sigma$ domain of $10^{10}<M_*/M_\odot<10^{11}$ galaxies predicted from the total \sfrIR. 
    All of the unobscured SF we measure could be attributed to low mass galaxies, which is consistent with the high contribution of this population to \sfrIR\ \citep{Alberts2021}, and their relatively low obscured fractions. 
    }
    \label{fig:sfrUV}
\end{figure}

Predictions using Equation \ref{eq:fobs} are shown as gray shaded regions on Figure \ref{fig:sfrUV}, and correspond to the total unobscured star-formation arising from low mass galaxies in our cluster stacks assuming field-like conditions. These values do not change if we adopt stellar mass functions derived from clusters because the 
stellar mass function of star-forming galaxies does not depend on environment \citep[e.g.,][]{vanderBurg2018}. From the uncertainties on the stellar mass functions, $f_{obs}(M_*)$, and the SFR calibrations, we estimate a systematic uncertainty on our predicted \sfrUV\ of 30\%. 
As shown in Figure~\ref{fig:sfrUV}, our cluster measurements of \sfrUV\ are broadly consistent with what is expected to arise from low mass $z\sim0.8-1.1$  field galaxies. At $z\sim0.57$, the total \sfrUV\ we measure is a factor of $2.5$ below the predictions, but within the $1\sigma$ uncertainties on both measured and predicted quantities. 
If significant, this deficit could reflect the rapid increase in quenching efficiency for $9<\log\mathrm{M/M_\odot}<10$ galaxies from $z\sim2$ to $z\sim0.5$ \citep{DeLucia2004,Stott2007,Kawinwanichakij2017} which is plausible given that such low mass galaxies are easily perturbed in clusters \citep{Boselli2008,Boselli2014}, and because they exhibit low obscured fractions \citep{Whitaker2017}.
At all redshifts, high mass galaxies have $f_{obs}\sim1$ and likely do not contribute significantly to \sfrUV. 

Adding this unobscured star-formation component to the previous measured obscured component reveals that the total specific star-formation rate in galaxy clusters from $z\sim0.57-1.1$ is declining below that of the field population (see Figure 12 in \alberts, see also \citealt{Alberts2014,Brodwin2013}). In other words 
the GALEX data did not reveal a significant missing component of star-formation from the IR analysis. This is consistent with the high environmental quenching efficiencies of low mass galaxies reported at $z<1.5$ \citep{Kawinwanichakij2017,Papovich2018,McNab2021}, which as a population must therefore play an important role in setting the evolution of clusters (\alberts, \citealt{vanderBurg2018}). 
There is tentative evidence for low gas reservoirs and low depletion timescales in $\log\mathrm{M/M_*}>9.5$ cluster star-forming galaxies at $z\sim0.75$ \citep{Betti2019}, and at $z\sim1.4$ (Alberts et al., submitted)
which supports that gas removal is driving a ``delayed then rapid'' quenching scenario \citep[i.e.,][]{Wetzel2013}; 
however, future observations of the gas content in lower mass cluster galaxies are necessary to further test this scenario. 

\subsection{Stripping of Dusty Envelopes in Galaxy Clusters at $z\sim0.57$}
As outlined in Section \ref{sec:nuv_spire}, we find evidence that the unobscured SF traced by the GALEX/$NUV$ data includes a centrally concentrated component not found for the dust-obscured SF measured by \textit{Herschel} in the $z\sim0.57$ cluster stack. One mechanism that could affect the obscured SF fraction at different radii is hydrodynamic gas stripping by the ICM (ram pressure stripping, \citealt{Boselli2021}), a common feature observed in infalling cluster galaxies at low- (e.g., \citealt{Poggianti2017,Longobardi2020}) and high-redshift \citep{Boselli2019,Noble2019}. As galaxies fall into the inner regions of clusters, they experience progressively more gas stripping in the outer disk regions, while simultaneously processing denser gas within the plane of the disk into stars. In particular, simulations show that most of a star-forming galaxy's halo gas is stripped before arriving at the virial radius of the host cluster, but that star-forming disks remain unperturbed until $<0.5R_{vir}$ \citep{Zinger2018}. These galaxies also tend to be preferentially found on radial orbits \citep{Lotz2019}. One could plausibly expect more unobscured SF in the inner projected 2D radial profiles where the averaged galaxy population has preferentially had dusty envelopes removed by interactions with the ICM. This could produce the centrally concentrated component of unobscured SF at $z\sim0.57$ but not at $z\sim0.86$ as galaxies have had more time to evolve towards the cluster centers at lower redshifts. 

\subsection{Stellar to halo mass ratios} 
The ratio of total stellar to halo mass in cluster members is representative of the efficiency of stellar mass assembly in dense environments, itself a function of competing feedback mechanisms on galaxy and halo scales. Star-formation and AGN feedback in galaxy/group and cluster scale halos, respectively, have been invoked to explain the small \mstar/\mhalo\ ratios which can be implemented in analytic and numerical models to test theories of galaxy formation within the framework of dark matter. By measuring the total cluster light from UV to IR wavelengths, we place constraint on empirical trends in \mstar/\mhalo\ for cluster galaxies which can be used to test theoretical predictions. 

\begin{figure}
    \centering
    \includegraphics[width=.48\textwidth]{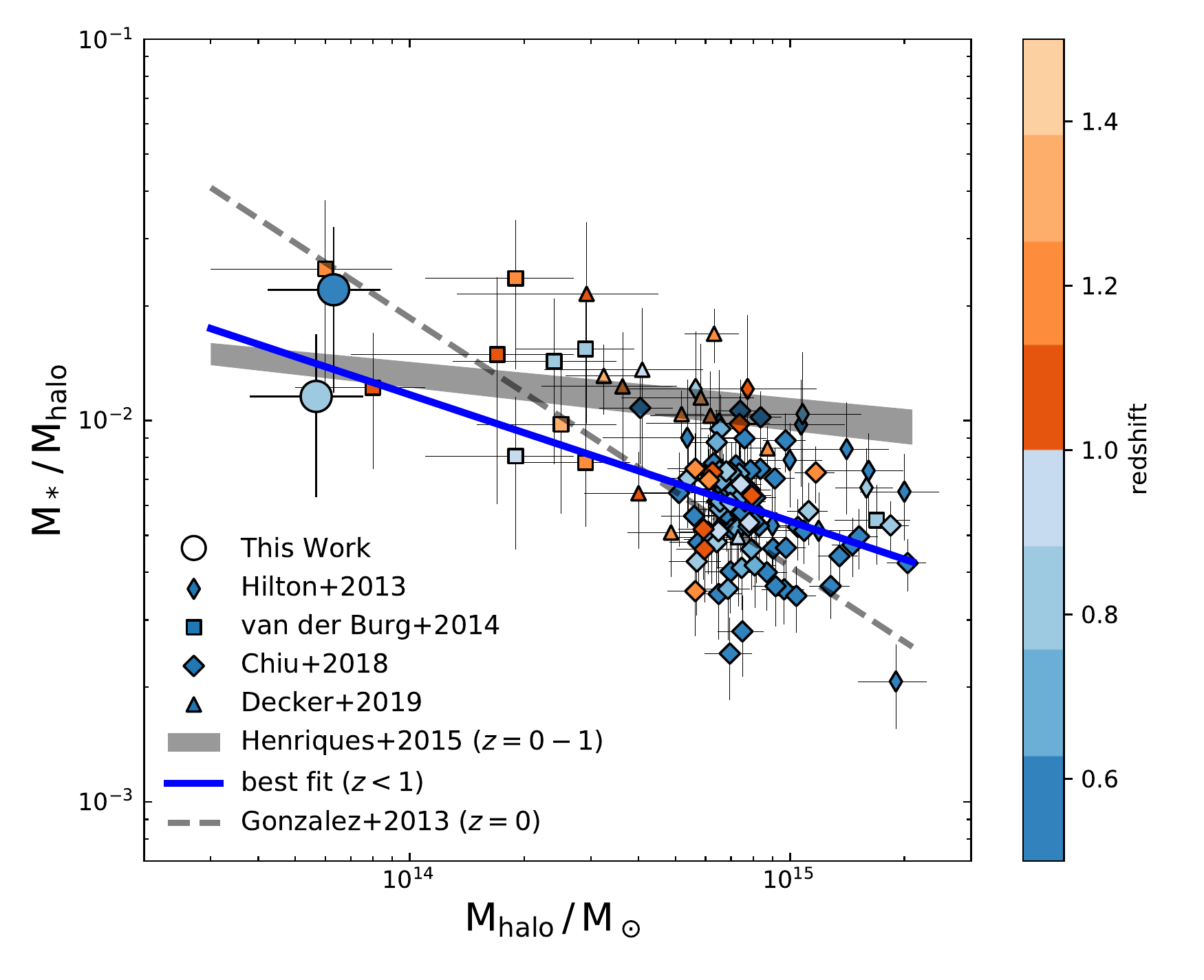}
    \caption{Stellar-to-halo mass fraction as a function of halo mass. This figure is updated from \alberts, using the new stellar mass estimates and uncertainties at $z<1$. We compare our observations against high$-z$ massive clusters in the literature \citep{Chiu2018,Hilton2013,Decker2019}, semi-analytic models of galaxy formation (shaded grey line, \citealt{Henriques2015}), and $z\sim0$ X-ray clusters (dashed grey line, \citealt{Gonzalez2013}). A fit to the $0<z<1$ data is shown in blue.  The data supports an anti-correlation between $\mathrm{M_*/M_{halo}}$ and $\mathrm{M_{halo}}$ steeper than what is predicted from models at all redshifts.  
    }
    \label{fig:mhalo}
\end{figure}

Clusters in the full ISCS sample have a mean halo mass of $\mathrm{\log M_h/M_\odot=13.8}$, independent of redshift \citep{Brodwin2007,Lin2013,Alberts2014}. In Figure \ref{fig:mhalo}, we show \mstar/\mhalo\ vs.~\mhalo\ in our $z<1$ stacked cluster sample, where the GALEX data improves constraint on \mstar\ (Table~\ref{tab:seds}), compared against observations of more massive cluster samples between $z\sim0-1.5$ \citep{Gonzalez2013,Hilton2013,vanDerBurg2014,Chiu2018,Decker2019}, and simulations \citep{Henriques2015}. Note that we compare against semi-analytical galaxy formation models as opposed to more common \mstar/\mhalo\ relations for central galaxies (e.g., \citealt{Moster2010}) because our method measures all of the light from the cluster. The \cite{Henriques2015} model predicts little correlation between \mstar/$\mathrm{M_{halo}}$ and $\mathrm{M_{halo}}$ at $z=0-1$, whereas empirical results from X-ray selected $z\sim0$ clusters suggest a sharp anti-correlation between the two quantities \citep{Gonzalez2013}. Results from our previous work (\alberts) could not distinguish between either scenario owing to large uncertainties on \mstar; however, the added GALEX/$NUV$ data improves constraint on derived stellar masses at $z<1$ (Table \ref{tab:seds}). We fit a trend line to our data and the $z<1$ literature, and find an anti-correlation between \mstar/$\mathrm{M_{halo}}$ and $\mathrm{M_{halo}}$. The slope of this trend (blue line, Figure~\ref{fig:mhalo}) is shallower than $z\sim0$ clusters \citep{Gonzalez2013}, but steeper than what is observed in the SAMs. Our results support a stellar-mass-dependent star-formation efficiency in clusters, with more massive systems exhibiting lower efficiencies. Further observations of $z\sim0.5-1$ clusters with  $\mathrm{M_{halo}<10^{14}\,M_\odot}$ will provide more detailed constraint on the redshift evolution. Nevertheless, our results stress the importance of placing strong constraints at rest-frame UV wavelengths to elucidate how star-formation proceeded in cluster environments. 

\subsection{Contaminants to the UV Emission}
The hot gas in clusters is a source of UV photons in addition to the emission produced by massive, hot stars. Electrons at the virial temperature of halos emit free-free radiation extending into the UV \citep{Sarazin1999,Sarazin2005}, which can also be contaminated by inverse Compton (IC) scattering of electrons, accelerated to relativistic velocities in shocked gas during mergers within the cluster environment. \cite{Welch2020} rule out both mechanisms as likely contributors to the UV emission detected in GALEX $FUV$ stacks of SZ-selected Planck cluster galaxies at $z<0.3$. Following their argument, we also rule out these sources of non-stellar UV photons as significant contributors to the measured flux densities in our cluster stacks: The typical spectral luminosity of free-free emission by halo electrons is $\sim10^{27}\,\mathrm{erg\,s^{-1}\,Hz^{-1}}$, and $\sim 10^{28}\,\mathrm{erg\,s^{-1}\,Hz^{-1}}$ for IC scattering off of shocked gas \citep{Sarazin2005}, corresponding to flux densities on the order of $10^{-5}$ mJy and $10^{-3}$ mJy respectively in the GALEX $NUV$ filter profiles at $z\sim0.5-1.4$. The flux densities of UV photons from IC scattering and free-free emission are 3-5 orders of magnitude less than the measured flux in our weighted mean cluster stacks.

Another source of contaminant flux is the UV upturn which manifests as a rise in the SED of quiescent galaxies at rest-frame UV wavelengths \citep{Burstein1988,OConnell1999}. The UV upturn is generally attributed to hot horizontal branch stars, and is less prevalent at earlier times when stars have had less time to evolve. For this reason, we expect the UV upturn to contribute negligibly to the stacked GALEX flux measured in our highest redshift bins \citep[e.g.,][]{Ali2018_1,Dantas2020}. 

To quantify the contribution from the UV upturn to the measured flux in our $z\sim0.57$ bin, we estimate the UV flux arising from horizontal branch stars in quiescent cluster members based on the typical UV-optical color of UV-upturn galaxies. At this redshift, the GALEX $NUV$ band corresponds roughly to the rest frame $FUV$ ($\sim$ 1100--1800 \AA\ at $z = 0.57$). We assume a rest-frame $FUV-r$ color of 5.5 for UV upturn galaxies \citep{Yi2011,Ali2018_1,Phillipps2020}. The colors of UV upturn galaxies are not expected to vary significantly with redshift or environment \citep{Boissier2018}. We also assume 
that $\lesssim25\%$ of cluster galaxies will exhibit a UV upturn  based on the fraction of early-type galaxies in a cluster \citep{Desai2009,Simard2009,Castellon2014,Lin2014,Jian2018} and the fraction of these displaying a UV upturn \citep{Ali2018_1,Dantas2020}. In reality, 
the fraction of cluster galaxies which are early-type/quiescent depends on the cluster stellar mass function, but assuming a constant value of $25\%$ will yield an upper limit on the UV flux from UV-upturn galaxies. From the best-fit $z\sim0.57$ SED, the $r$  band ($\approx$6500 \AA) flux within a 1 Mpc radius aperture is $1.05$ mJy. The maximum potential contribution to the total UV flux from the UV upturn is therefore $\approx25\%\times10^{-0.4\cdot 5.5}\times1.05$ mJy, corresponding to  $\approx$ 1.5 $\mu$Jy.
While there may be some contribution from the UV upturn to our GALEX measurements, it is $\lesssim15\%$ in the measured cumulative fluxes of our lowest redshift bin at most and thus is not a dominant factor. Assuming the total stellar mass (and hence the UV upturn $r$-band flux) scales with the WISE radial profiles, the contribution of the UV upturn to the flux within 200 kpc is at most 20\% of the total UV flux in this region where early-type galaxies dominate \citep{Dressler1997}, which is insufficient to explain the central excess in the z$\sim$ 0.57 UV radial profile.

\section{Conclusion\label{sec:conclusion}}
In this work, we stack GALEX/$NUV$ maps of galaxy clusters between $z\sim0.5-1.6$ to constrain the total UV emission from cluster-member galaxies, including those which go undetected in individual observations. We measure aperture-integrated photometry and radial flux profiles from the stacked maps in four redshift bins, $z=0.5-0.7,\,0.7-1.0,\,1.0-1.4$ and $z=1.4-1.6$, which we combine with near- to far-IR stacks of the same clusters from \citealt{Alberts2021} (\alberts). From fits to the UV through IR SEDs, we measure total stellar masses, and the ratio of obscured to total star-formation in clusters ($f_{obs}$) over a broad range in redshift. We stress that this method includes the light from lower mass galaxies expected to harbour most of the unobscured star-formation. Such galaxies are commonly missed by studies of bright cluster members. Our main conclusions are summarized as follows:  
\begin{enumerate}
    \item The UV flux implies low amounts of unobscured star-formation relative to the dust-obscured star-formation measured in the IR ($f_{obs}>0.8$). Previous work found the IR light to be dominated by $\log\mathrm{M_*/M_\odot}<10$ galaxies, which also dominate the UV and therefore the total star-formation in clusters. The predicted \sfrUV\ from this low-mass population could explain all of the unobscured SF we measure at $z<1$ if low-mass galaxies have as much dust-obscured star-formation as their field counterparts. 
    \item Adding in the UV confirms that the total (s)SFR in galaxy clusters follows a declining trend from $z\sim1$ to $z\sim0.5$ relative to the field. 
   Because low mass galaxies dominate both the IR and UV, and therefore dominate the total star-formation in $z\sim0.5-1.4$ clusters, environmental quenching of low-mass galaxies must be an important driver of cluster evolution. 
    \item The UV radial profile of $z\sim0.57$ clusters have a centrally concentrated component not found in the IR radial profile. This could arise from gas stripping via interactions between infalling cluster galaxies and the intracluster medium, which can strip the obscuring material from the outer envelopes of star-forming galaxies as they make their way into the cluster center. 
    \item Improved stellar mass estimates favor an anti-correlation between \mstar/\mhalo\ and \mhalo, consistent with trends found in $z\sim0$ clusters \citep{Gonzalez2013}. 
    This is in conflict with some semi-analytic models that predict a flatter stellar-to-halo mass function of \mhalo. 
\end{enumerate}
Total light stacking of galaxy clusters allows for the full accounting of galaxies in wavelength and redshift regimes that are otherwise limited by observational depth. In the future, this technique may be applied to other wavelengths, such as the millimeter regime using the upcoming TolTEC camera on the Large Millimeter Telescope in order to measure the evolution of the total dust mass in clusters. With other large area samples that have different mass selection functions (e.g., MaDCoWS; \citealt{Gonzalez2019}), this technique may also be used to test for halo mass-dependent evolution in the panchromatic SED of galaxy clusters out to high$-z$ and place further constraint on models for galaxy formation within the context of large scale structure.

\begin{acknowledgments}
We thank the referee for their insightful comments that strengthened this work. JM and AP thank M. Weinberg for helpful discussion on non-parametric statistical tests. The authors acknowledge financial support from NASA through the Astrophysics Data Analysis Program, grant number 80NSSC19K0582. GALEX is a NASA Small Explorer, and we gratefully acknowledge NASA’s support for construction, operation, and science analysis for the GALEX mission, developed in cooperation with the Centre National d’Etudes Spatiales of France and the Korean Ministry of Science and Technology. 
\end{acknowledgments}

\bibliography{references}

\end{document}